\documentclass[iop,apj,tighten,twocolappendix,numberedappendix]{emulateapj}
\usepackage{apjfonts}

\usepackage{graphicx}
\usepackage{amsmath}
\usepackage{amssymb}
\usepackage{color}
\usepackage{mathtools}
\usepackage{url}

\usepackage[breaklinks,colorlinks,citecolor=blue,linkcolor=red]{hyperref} 
\usepackage[all]{hypcap}

\newcommand\msun{\, \rm M_\odot}

\newcommand\kms{\, \rm km\,s^{-1}}

\newcommand\gpcyr{\, \rm Gpc^{-3}\,yr^{-1}}

\newcommand\vk{{v_\mathrm{k}}}

\begin{document}

\title{Impact of natal kicks on merger rates and spin-orbit misalignments of black hole -- neutron star mergers}

\author{Giacomo Fragione\altaffilmark{1,2}, Abraham Loeb \altaffilmark{3}, Frederic A.\ Rasio\altaffilmark{1,2}}
 \affil{$^1$Center for Interdisciplinary Exploration \& Research in Astrophysics (CIERA), Evanston, IL 60202, USA} 
  \affil{$^2$Department of Physics \& Astronomy, Northwestern University, Evanston, IL 60202, USA}
  \affil{$^3$Astronomy Department, Harvard University, 60 Garden St., Cambridge, MA 02138, USA}

\begin{abstract}
The long wait for the detection of merging black hole -- neutron star (BH--NS) binaries is finally over with the announcement by the LIGO/Virgo/Kagra collaboration of GW200105 and GW200115. Remarkably, the primary of GW200115 has a negative spin projection onto the orbital angular momentum, with about $90\%$ probability. Merging BH--NS binaries are expected to form mainly through the evolution of massive binary stars in the field, since their dynamical formation in dense star clusters is strongly suppressed by mass segregation. In this paper, we carry out a systematic statistical study of the binary stars that evolve to form a BH--NS binary, considering different metallicities and taking into account the uncertainties on the natal kick distributions for BHs and NSs and on the common envelope phase of binary evolution. Under the assumption that the initial stellar spins are aligned with the binary angular momentum, we show that both large natal kicks for NSs ($\gtrsim 150\kms$) and high efficiencies for common envelope ejection are required to simultaneously explain the inferred high merger rates and the large spin-orbit misalignment of GW200115.
\end{abstract}

\section{Introduction}
\label{sect:intro}

Gravitational wave (GW) detectors promise to observe hundreds of merging black holes (BHs) and (NSs) in the next few years. The second Gravitational Wave Transient Catalog (GWTC-2) by the LIGO/Virgo/Kagra (LVK) Collaboration, which includes events through the first half of the third observational run \citep{lvc2020cat}, lists tens of BH--BH mergers and two NS--NS mergers \citep{abbott2017gw170817,abb2020}. The first mergers of BH--NS binaries, GW200105 and GW20115, have only been confirmed recently \citep{AbbottAbbott2021}. The corresponding BH--NS merger rate density is $45^{+75}_{-33}\gpcyr$, if GW200105 and GW200115 are representative of the entire BH--NS binary population, or $130^{+112}_{-69}\gpcyr$ under the assumption of a broader distribution of component masses.

BH--NS mergers are particularly important because they can produce an electromagnetic (EM) counterpart. The evolution of merging BH--NS binaries spans three phases: the inspiral ($\gtrsim 10^6$\,yr) due to GW emission, the merger phase ($\sim 1$\,ms), which can result in either the tidal disruption of the NS or its plunge into the BH, and, for disrupting systems only, a post-merger phase ($\sim 1$\,s) where NS matter debris can be ejected or accreted onto the BH. Mergers that produce an EM counterpart can provide crucial constraints on the BH accretion process and unique information on the nuclear equation of state, NS crust physics, and NS magnetospheres \citep[e.g.,][]{Pannarale2011,Foucart2012,tsang2012,FoucartHinderer2018,AscenziDeLillo2019,HindererNissanke2019,CoughlinDietrich2020,FragioneLoeb2021}. Moreover, BH--NS mergers have been invoked to explain the formation of r-process elements, to improve tests of fundamental physics, and to constrain models of ultra-relativistic jets \citep[e.g.,][]{KyutokuIoka2014,HotokezakaBeniamini2018,ThraneOslowski2020,SilvaHolgado2021}.

Despite these first detections, the origin of BH--NS mergers remains highly uncertain. BH--NS mergers can be produced as a result of the evolution of field binaries, with predicted merger rates consistent with the empirical LVK estimate within the large uncertainties of binary stellar evolution models \citep[e.g.,][]{demink2016,gm2018,kruc2018,BroekgaardenBerger2021,ShaoLi2021}. In dense star clusters, where stellar BHs dominate the cluster cores and prevent the mass segregation of NSs \citep{frag2018,ye2019millisecond,Kremer2019d}, the rates of dynamical interactions involving NSs are reduced and merging BH--NS cannot be efficiently formed. Only after most of the stellar BHs have been ejected from a cluster can the NSs then efficiently segregate into the innermost core and possibly form binaries that later merge. A number of papers have estimated the BH--NS merger rate from dense star clusters to be several orders of magnitude smaller than LVK estimated rates \citep{clausen2013black,bae2014compact,belczynski2018origin,FragioneBanerjee2020,asedda2020,ye2020}. 
Stellar dynamics can still play a role in shaping the progenitor massive binaries born in young star clusters, with corresponding merger rates matching the LVK rates, when favorable initial conditions are assumed and fractal initial configurations are taken into account \citep{rastello2020}. \citet{frl2019a,frl2019b} have proposed that BH--NS mergers can be a natural outcome of the evolution of triple star systems that undergo Lidov-Kozai cycles. However, merger rates consistent with the LVK results can only be achieved if very low natal kicks for BHs and NSs are assumed. Finally, dynamical assembly in AGN disks could also produce higher rates, even though there are still major uncertainties in the models \citep{YangGayathri2020,TagawaKocsis2021}. Clearly, the observed properties of BH--NS mergers provide critical information to help us determine which of these processes contributes the most to the formation and mergers of compact object binaries.

\begin{figure} 
\centering
\includegraphics[scale=0.595]{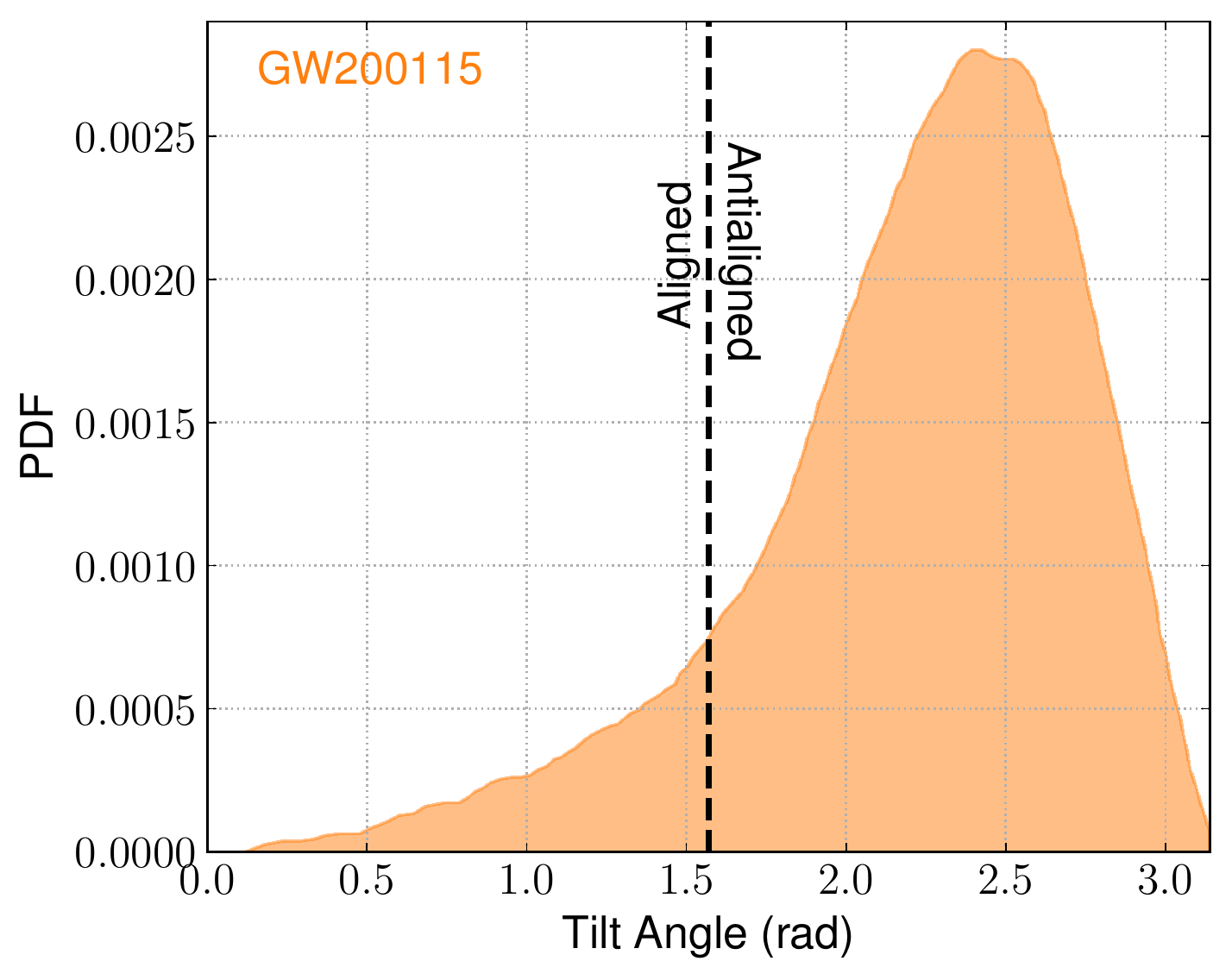}
\caption{Probability distribution function (PDF) of the misalignment of the primary spin in GW200115 with respect to the orbital angular momentum.}
\label{fig:tiltligo}
\end{figure}

One of the most interesting results from the first two BH--NS events concerns the orientation of their component spins \citep{AbbottAbbott2021}. While the orientation of the primary spin for GW200105 is unconstrained, the primary of GW200115 has a negative spin projection onto the orbital angular momentum at $88\%$ probability, with the spin-orbit angle estimated to be $2.30^{+0.59}_{-1.18}$ rad (see Figure~\ref{fig:tiltligo}). If BH--NS mergers are produced from field binaries, it is well known that the natal kicks imparted to the BH and NS can significantly tilt the orbital and spin angular momenta \citep[e.g.,][]{Kalogera2000}. In this Letter, we study the role of the recoil kick magnitude in the formation of BH--NS binaries from field binaries. Under the assumption of aligned stellar spins and orbital angular momentum in the progenitor binaries, we show that only large natal kicks and high efficiencies for common envelope ejection can explain the observations.

This Letter is organized as follows. In Section~\ref{sect:models} we describe our models and assumptions. In Section~\ref{sect:res} we discuss the expected distributions of delay times, merger rates, and misalignment angles for merging BH--NS binaries. Finally, in Section \ref{sect:conc}, we summarize our findings and draw our conclusions.

\section{Method}
\label{sect:models}

In all our models, we sample the initial mass $m_1$ of the primary from a \citet{kroupa2001} initial mass function,
\begin{equation}
\frac{\mathrm{d}N}{\mathrm{d}m} \propto m^{-2.3},
\label{eqn:massfunc}
\end{equation}
in the mass range $[20\,\msun$--$150\,\msun]$, appropriate for BH progenitors. We adopt a flat mass ratio distribution to determine the initial secondary mass ($m_2$), consistent with observations of massive binary stars \citep{sana12,duch2013,Sana2017}. The distributions of the orbital periods (in days) are extracted from
\begin{equation}
f(\log_{10} P) \propto (\log_{10} P)^{-0.55}
\end{equation}
in the range $[0.15$--$5.5]$ \citep{sana12}, while we adopt a thermal distribution for the eccentricity.

We then evolve the binaries using the stellar evolution code \textsc{bse} \citep{hurley2000comprehensive,hurley2002evolution}. We use the latest version of \textsc{bse} from \citet{BanerjeeBelczynski2020}, updated with the most up-to-date prescriptions for stellar winds and remnant formation. Importantly, the current version includes the most recent theoretical results on pulsational pair instabilities, which limit the maximum BH mass to about $50\,\msun$ \citep{bel2016b}, and produce remnant populations consistent with those from \textsc{StarTrack} \citep{belc2008}.  

When stars evolve to form a compact object, they may receive a natal kick due to recoil from an asymmetric supernova (SN) explosion. We assume that the velocity kick magnitude obeys a Maxwellian distribution,
\begin{equation}
p(v_\mathrm{kick})\propto v_\mathrm{kick}^2 \exp \left ( -\frac{v_\mathrm{kick}^2}{2\sigma^2} \right ),
\label{eqn:vkick}
\end{equation}
with velocity dispersion $\sigma$. This quantity is highly uncertain. For example, $\sigma=265$ km s$^{-1}$ was inferred from the proper motions of pulsars by \citet{hobbs2005}. However, \citet{arz2002} found a bimodal distribution with characteristic velocities of $90\kms$ and $500\kms$ based on the velocities of isolated radio pulsars, while \citet{BeniaminiPiran2016} found evidence for a low-kick population ($\lesssim 30\kms$) and a high-kick population ($\gtrsim 400\kms$) based on observed binary NSs. Finally, NSs born as a result of the electron-capture SN process could receive no kick at birth or only a very small one due to an asymmetry in neutrino emission \citep{pod2004}. For BHs, we adopt natal kicks from Eq.~\ref{eqn:vkick} as for NSs, but with  $\sigma$ scaled down linearly with increasing mass fallback fraction \citep{RepettoDavies2012,Janka2013}. For details see \citet{BanerjeeBelczynski2020}.

In order to compute the spin-orbit misalignment produced as a result of natal kicks, we extract them from \textsc{bse} and compute the tilt of the binary orbit (whenever the orbit remains bound). For example, let us assume that the SN event happens first in $m_1$ in a binary of initial semi-major axis $a$ and eccentricity $e$. Before the SN takes place, energy conservation implies
\begin{equation}
|{\bf{v}}|^2=\mu\left(\frac{2}{r}-\frac{1}{a}\right)\,,
\label{eqn:vcons}
\end{equation}
where $\bf{v}$ is the binary relative velocity, $\mu=G(m_1+m_2)$, and $\bf{r}$ is the binary relative separation, while the angular momentum $\bf{h}$ is related to the orbital parameters by
\begin{equation}
|{\bf{h}}|^2=|{\bf{r}}\times {\bf{v}}|^2=\mu a(1-e)^2\,.
\label{eqn:hcons}
\end{equation}
After the SN event, the orbital semi-major axis and eccentricity change due to the mass loss $\Delta m$ in $m_1$ and the natal kick $\bf{v_k}$ (assumed to be isotropic). The new total mass of the binary decreases to $m_{\rm in,n}=m_{1,n}+m_2$, where $m_{1,n}=m_1-\Delta m$, while the new relative velocity becomes ${\bf{v_n}}={\bf{v}}+{\bf{\vk}}$. Assuming that the SN takes place instantaneously, that is ${\bf{r_\mathrm{n}}}={\bf{r}}$, the misalignment between the post-SN and pre-SN orbit, $\Delta \theta_1$, can be obtained from 
\begin{equation}
\Delta \theta_1=\arccos \left(\frac{\bf{h}\cdot \bf{h_{\rm n}}}{h\ h_{\rm n}} \right) \,,
\end{equation}
where ${\bf{h_n}}={\bf{r}}\times {\bf{v_n}}$ is the post-SN angular momentum. The misalignment produced as a result of the SN explosion of $m_2$, $\Delta \theta_2$, can be computed similarly. Under the assumption that the initial stellar spins are aligned with the binary angular momentum, $\Delta \theta=\Delta \theta_1+\Delta \theta_2$ corresponds to the final tilt between the spins of the binary components and the binary angular momentum.

We consider in total 4 values of $\sigma$, namely $10\kms$, $40\kms$, $150\kms$, $260\kms$, and, for each value of $\sigma$, stellar evolution with 6 different assumed metallicities, namely $Z=0.0002$, $0.001$, $0.002$, $0.005$, $0.01$, $0.02$. We also consider three different values of the common-envelope energy efficiency parameter, $\alpha=1$, $3$, $5$ \citep{hurley2002evolution}. For each combination, we compute 35,000~evolutions, for a total of about 2.5~million simulated binaries.

\section{Results}
\label{sect:res}

\begin{figure} 
\centering
\includegraphics[scale=0.585]{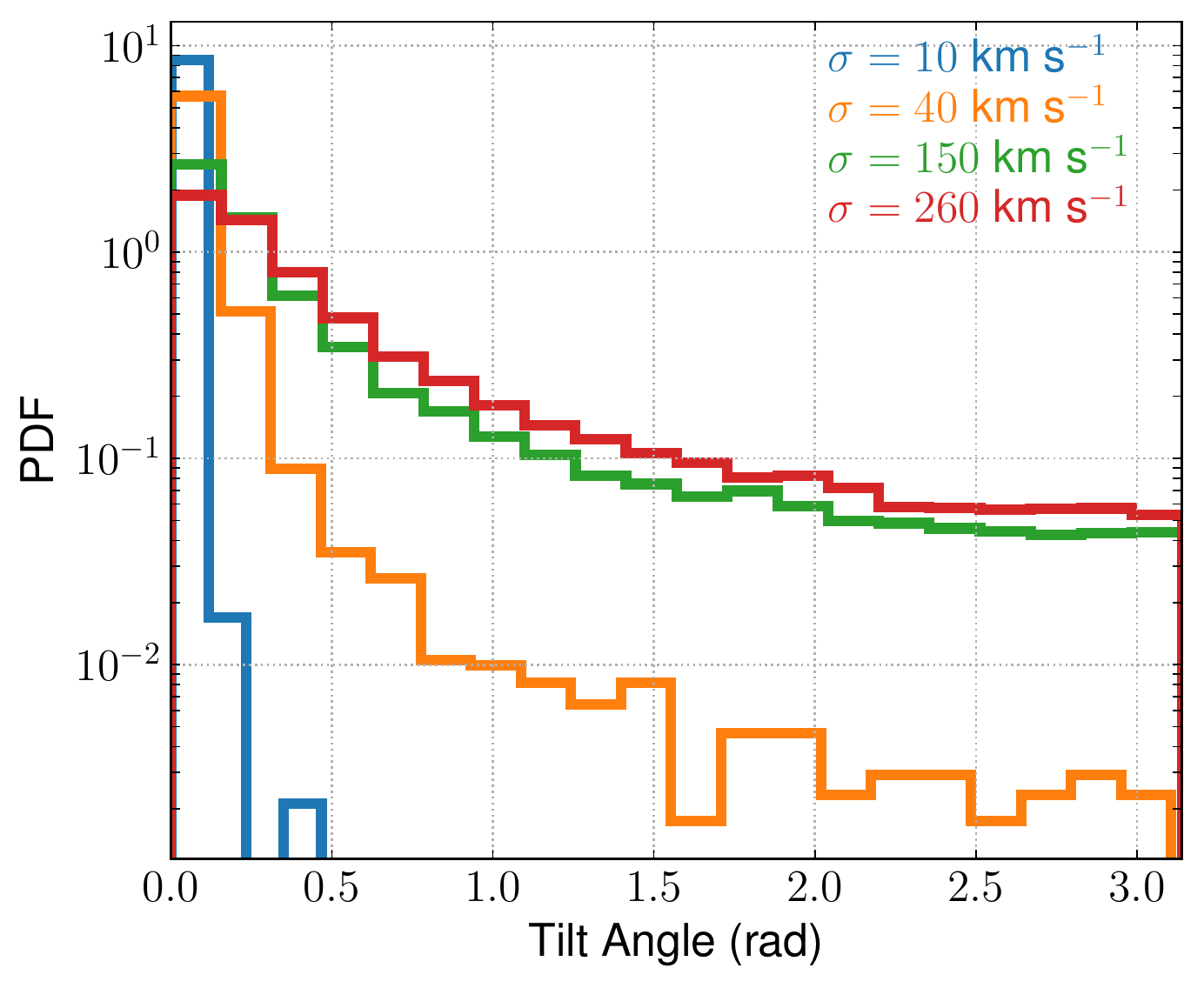}
\caption{Distribution of tilt angles resulting from natal kicks for merging BH--NS binaries, using all metallicities and $\alpha=1$. Different colors represent different values of the natal-kick velocity dispersion $\sigma$ (see Eq.~\ref{eqn:vkick}).}
\label{fig:tiltsim}
\end{figure}

Figure~\ref{fig:tiltsim} shows the distribution of tilt angles as a result of natal kicks for merging BH--NS binaries (all the systems with delay time less than 13.8~Gyr) as a function of $\sigma$, for all metallicities and $\alpha=1$. As expected, we find that larger natal kicks produce more mergers with large tilt angles. Moreover, there are no merging BH--NS systems with tilt angles $> 0.5$~rad when $\sigma=10\kms$. For $\sigma=40\kms$, tilt angles can be as large as about $3$~rad, but they are nearly 100~times less likely than small misalignment angles ($\lesssim 0.5$~rad). For $\sigma=150\kms$ and $\sigma=260\kms$, this probability becomes about one order of magnitude larger. When assuming $\alpha=3$ and $\alpha=5$, there are no significant differences. Therefore, a GW200115-like event (see Figure~\ref{fig:tiltligo}) is more likely to be produced with high natal kicks.

To understand the distribution of tilt angles for a cosmologically-motivated population, we compute the differential merger rate of BH--NS binaries. We start with computing the merger rate as \citep[see also][]{SantoliquidoMapelli2020}
\begin{eqnarray}
R(z) &=& f_{\rm b} f_{\rm IMF} \frac{d}{dt_{\rm lb}(z)} \int^{z_{\max}}_{z} \Psi(\zeta) \frac{dt_{\rm lb}(\zeta)}{d\zeta} d\zeta \nonumber\\
& \times & \int^{Z_{\max}(\zeta)}_{Z_{\min}(\zeta)} \Phi(z, Z) \Pi(\zeta, Z) dZ\,,
\label{eqn:ratez}
\end{eqnarray}
where $f_{\rm b}=0.5$ is the fraction of stars in binaries \citep[e.g.,][]{ragh10,duch2013,Sana2017}, $f_{\rm IMF}=0.115$ is a correction factor that accounts for our truncation of the primary mass distribution $\ge 20\msun$ (assuming a \citet{kroupa2001} initial mass function), $t_{\rm lb}$ is the look-back time at redshift $z$ \footnote{For our calculations we assume the cosmological parameters from Planck 2015 \citep{PlanckCollaborationAde2016}.}. In Eq.~\ref{eqn:ratez}, $\Psi$ is the cosmic star formation history \citep{MadauDickinson2014}
\begin{equation}
\Psi(z)=0.01 \frac{(1+z)^{2.6}} {1.0 + [(1.0 + z) / 3.2]^{6.2}}\,{\rm M}_\sun\,{\rm yr}^{-1}\,{\rm Mpc}^{-3}\, ,
\label{eqn:madau}
\end{equation}
$\Phi$ is the merger efficiency at a given metallicity
\begin{equation}
\Phi(z, Z) = \frac{N_{\rm merger} (z, Z)}{M_{\rm tot} \, (Z)} \, ,
\end{equation}
where $M_{\rm tot} (Z)$ is the total simulated mass at metallicity $Z$ and $N_{\rm merger} (z, Z)$ the total number of BH--NS mergers at redshift $z$ originating from progenitors at metallicity $Z$, and $\Pi$ is the metallicity distribution at a given redshift, which we assume is described by a log-normal distribution with mean given by \citep{MadauFragos2017}
\begin{equation}
\log \langle Z/{\rm Z}_\odot \rangle = 0.153 - 0.074 z^{1.34}
\end{equation}
and a standard deviation of 0.5~dex \citep{DvorkinSilk2015}. 

\begin{figure} 
\centering
\includegraphics[scale=0.585]{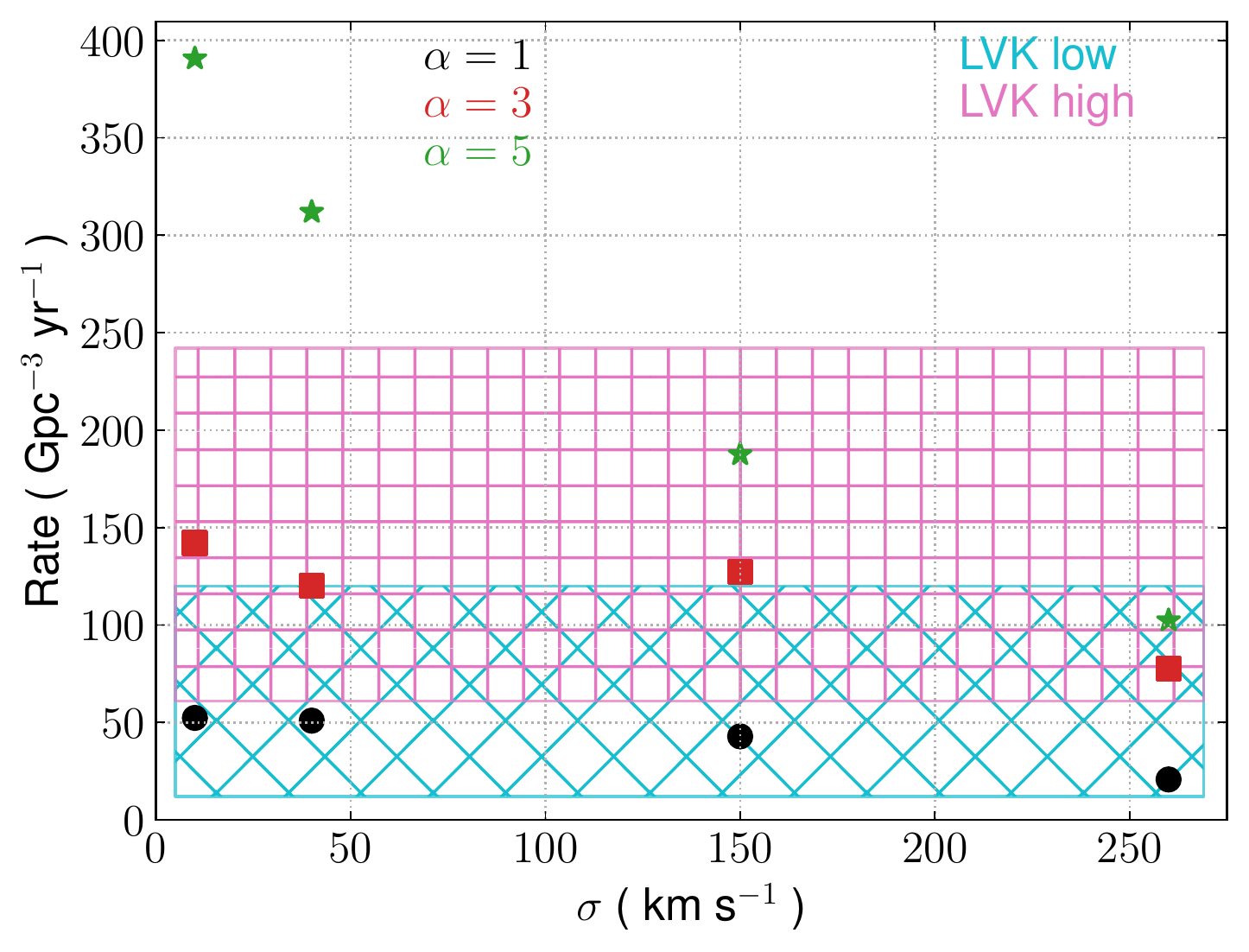}
\caption{Local merger rate density of BH--NS systems from field binaries as a function of the natal-kick velocity dispersion $\sigma$. Different colors represent different values of $\alpha$, which quantifies the energy available to unbind the envelope during common-envelope phases of binary evolution. The shaded regions represent the BH--NS merger rate estimated by LVK when assuming that GW200105 and GW200115 are representative of the NS--BH population (cyan; LVK low) and under the assumption of a broader distribution of component masses (pink; LVK high).}
\label{fig:rate}
\end{figure}

Figure~\ref{fig:rate} shows the local merger rate density of BH--NS systems as a function of $\sigma$ (see Eq.~\ref{eqn:vkick}) and for different assumed values of $\alpha$. We also show the BH--NS merger rate estimated empirically by LVK. We see that for $\sigma \le 150\kms$, BH--NS merger rates are about $50\gpcyr$ and $150\gpcyr$ for $\alpha=1$ and $\alpha=3$, respectively, about twice what we get for $\sigma=260\kms$. When $\alpha=5$, we estimate that the merger rates are about $400\gpcyr$, $300\gpcyr$, $200\gpcyr$, $100\gpcyr$ for $\sigma=10\kms$, $\sigma=40\kms$, $\sigma=150\kms$, $\sigma=260\kms$, respectively. While our inferred merger rates are in agreement with the LVK collaboration's lower estimate, they can account for their higher estimate only if $\alpha=3$ or $\alpha=5$, regardless of the assumed value for $\sigma$. When assuming $\sigma \le 40\kms$ and $\alpha=5$, our predicted merger rates become too high.

\begin{figure} 
\centering
\includegraphics[scale=0.61]{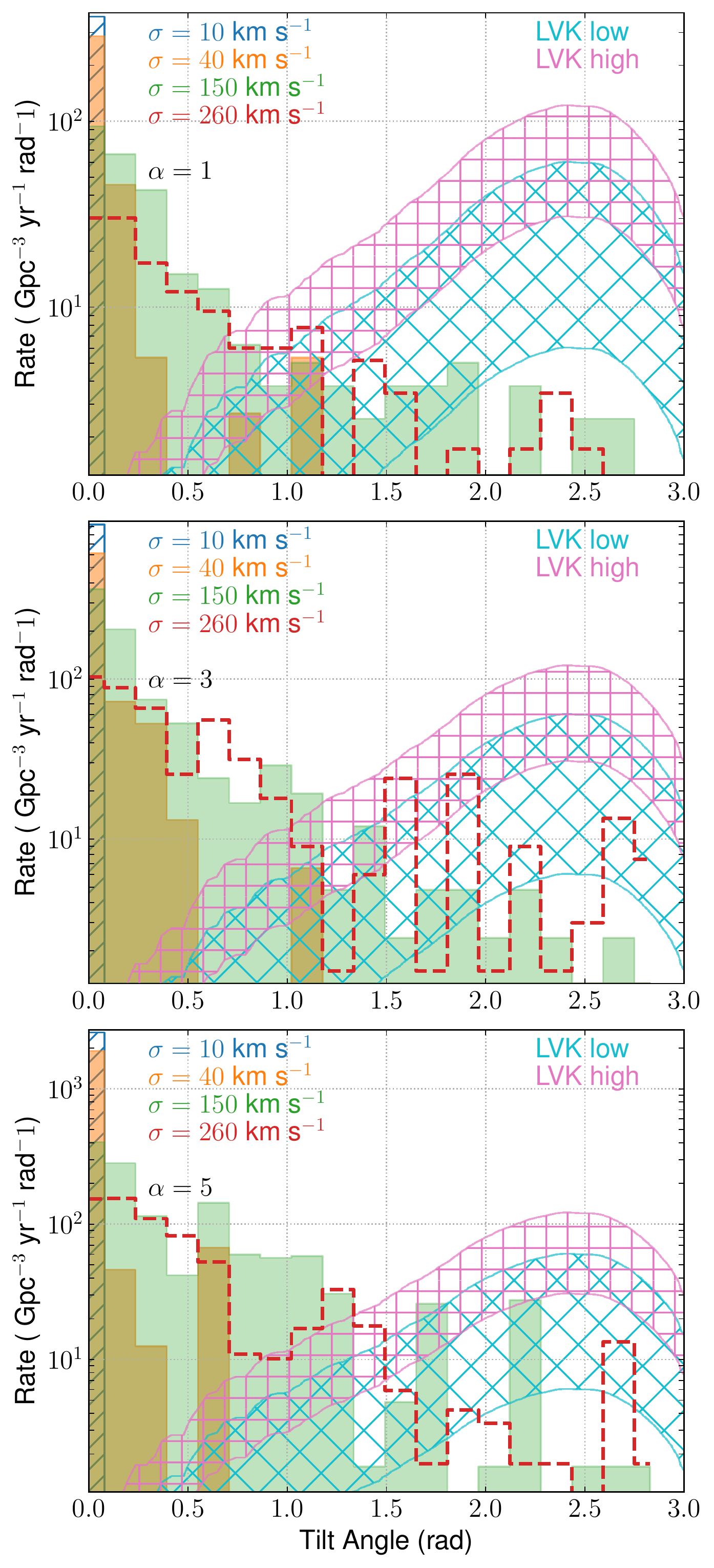}
\caption{Local differential merger rate density of BH--NS systems as a function of the tilt angle. Different colors represent different values of the natal-kick velocity dispersion $\sigma$. For reference, we show the two different LVK BH--NS merger rates (LVK low in cyan and LVK high in pink) weighted by the distribution of the misalignment of the primary spin in GW200115. Different panels represent different values of the common-envelope efficiency parameter $\alpha$. Top: $\alpha=1$; center: $\alpha=3$; bottom: $\alpha=5$.}
\label{fig:ratetilt135}
\end{figure}

To understand how the merger rate varies with tilt angle, we compute the differential merger rate as a function of the tilt angle as \citep[see also][]{Banerjee2021, Banerjee2021b}
\begin{equation}
\frac{dR(z, \theta)}{d\theta} = R(z)\ \Xi (\theta)\,,
\end{equation}
where
\begin{equation}
\Xi (\theta) = \frac{1}{N_{\rm tot}(z)}\frac{\partial N(z)}{\partial \theta}\,,
\end{equation}
defined such that $\int_0^\pi \Xi (\theta) d\theta=1$.
Figure~\ref{fig:ratetilt135} shows this rate density as a function of $\theta$. For reference, we also show the two different LVK empirical merger rates, which we weigh at different tilt angles by using the distribution of spin-orbit misalignments inferred for GW200115 (see Figure~\ref{fig:tiltligo}). The cases with low assumed values of $\sigma$ ($10\kms$ and $40\kms$) produce larger rates since fewer binaries are disrupted as a result of SN kicks, but they cannot account for large misalignment angles in these systems. On the other hand, the cases with high assumed values of $\sigma$ ($150\kms$ and $260\kms$) can explain large spin-orbit misalignments. However, high efficiencies of common-envelope ejection ($\alpha=3$ and $\alpha=5$) must be assumed in order to explain the empirical LVK merger rates.

\section{Summary and Conclusions}
\label{sect:conc}

The recent detection of GW200105 and GW20115, the first two BH--NS merger events, by the LVK collaboration provided the first empirical estimate of the BH--NS merger rate \citep{AbbottAbbott2021}. A key open question remains the astrophysical origin of these merging BH--NS binaries. While their formation is suppressed in dense star clusters as a result of the strong dynamical heating by BHs, massive binaries in the field can produce BH--NS mergers as the end point of isolated stellar evolution at a rate consistent with LVK estimates. Apart from matching the rate, the question we addressed here is whether isolated binary evolution can also reproduce the finding that the primary in GW200115 has a negative spin projection ($2.30^{+0.59}_{-1.18}$ rad) onto the orbital angular momentum, indicating a large spin-orbit misalignment. Here we carried out a broad statistical study of the field binary stars that evolve to form merging BH--NS binaries. We considered different metallicities and we took into account the significant uncertainties on the efficiency of common envelope ejection and natal kick magnitudes for compact objects. 

We found that the detection of a GW200115-like event indicates that large natal kicks ($\sigma \ge 150\kms$) are necessary at NS formation. We computed the differential merger rate of BH--NS binaries as a function of the spin-orbit misalignment angle and showed that only large natal kicks combined with high efficiencies for common envelope ejection can explain simultaneously the high empirical merger rate and the large spin-orbit misalignment detected for GW200115.

Future LVK observing runs promise to detect hundreds of merging BH--NS systems, thus providing a sufficiently high number to constrain much more tightly both the merger rate and spin-orbit misalignments as a function of redshift. In turn, through theoretical analyses like the one we sketched here, these observations will provide much deeper insights into the astrophysics of compact object formation and binary star evolution.

\section*{Acknowledgements}

GF is grateful to Sambaran Banerjee for useful discussions on stellar evolution and for updating \textsc{bse}. GF and FAR acknowledge support from NSF Grants~AST-1716762 and~AST-2108624 at Northwestern University. This work was also supported in part by Harvard's Black Hole Initiative, which is funded by grants from JFT and GBMF.

\bibliographystyle{yahapj}
\bibliography{refs}

\begin{thebibliography}{}
\providecommand\natexlab[1]{#1}
\providecommand\JournalTitle[1]{#1}

\bibitem[{Abbott {et~al.}(2017)Abbott, Abbott, Abbott, Acernese, Ackley, Adams,
  Adams, Addesso, Adhikari, Adya, {et~al.}}]{abbott2017gw170817}
Abbott, B.~P., Abbott, R., Abbott, T., {et~al.} 2017, \JournalTitle{\prl}, 119,
  161101

\bibitem[{{Abbott} {et~al.}(2020{\natexlab{a}})}]{abb2020}
{Abbott}, B.~P., {et~al.} 2020{\natexlab{a}},
  \href{http://dx.doi.org/10.3847/2041-8213/ab75f5}{\JournalTitle{\apjl}, 892,
  L3}

\bibitem[{{Abbott} {et~al.}(2021){Abbott}, {Abbott}, {Abraham}, {Acernese}, \&
  et~al.}]{AbbottAbbott2021}
{Abbott}, R., {Abbott}, T.~D., {Abraham}, S., {Acernese}, F., \& et~al. 2021,
  \href{http://dx.doi.org/10.3847/2041-8213/ac082e}{\JournalTitle{\apjl}, 915,
  L5}

\bibitem[{{Abbott} {et~al.}(2020{\natexlab{b}})}]{lvc2020cat}
{Abbott}, R., {et~al.} 2020{\natexlab{b}}, \JournalTitle{arXiv e-prints},
  arXiv:2010.14527

\bibitem[{{Arca Sedda}(2020)}]{asedda2020}
{Arca Sedda}, M. 2020,
  \href{http://dx.doi.org/10.1038/s42005-020-0310-x}{\JournalTitle{Communications
  Physics}, 3, 43}

\bibitem[{{Arzoumanian} {et~al.}(2002){Arzoumanian}, {Chernoff}, \&
  {Cordes}}]{arz2002}
{Arzoumanian}, Z., {Chernoff}, D.~F., \& {Cordes}, J.~M. 2002,
  \href{http://dx.doi.org/10.1086/338805}{\JournalTitle{\apj}, 568, 289}

\bibitem[{{Ascenzi} {et~al.}(2019){Ascenzi}, {De Lillo}, {Haster}, {Ohme}, \&
  et~al.}]{AscenziDeLillo2019}
{Ascenzi}, S., {De Lillo}, N., {Haster}, C.-J., {Ohme}, F., \& et~al. 2019,
  \href{http://dx.doi.org/10.3847/1538-4357/ab1b15}{\JournalTitle{\apj}, 877,
  94}

\bibitem[{Bae {et~al.}(2014)Bae, Kim, \& Lee}]{bae2014compact}
Bae, Y.-B., Kim, C., \& Lee, H.~M. 2014, \JournalTitle{\mnras}, 440, 2714

\bibitem[{{Banerjee}(2021{\natexlab{a}})}]{Banerjee2021}
{Banerjee}, S. 2021{\natexlab{a}},
  \href{http://dx.doi.org/10.1093/mnras/staa2392}{\JournalTitle{\mnras}, 500,
  3002}

\bibitem[{{Banerjee}(2021{\natexlab{b}})}]{Banerjee2021b}
---. 2021{\natexlab{b}},
  \href{http://dx.doi.org/10.1093/mnras/stab591}{\JournalTitle{\mnras}, 503,
  3371}

\bibitem[{{Banerjee} {et~al.}(2020){Banerjee}, {Belczynski}, {Fryer},
  {Berczik}, \& et~al.}]{BanerjeeBelczynski2020}
{Banerjee}, S., {Belczynski}, K., {Fryer}, C.~L., {Berczik}, P., \& et~al.
  2020,
  \href{http://dx.doi.org/10.1051/0004-6361/201935332}{\JournalTitle{\aap},
  639, A41}

\bibitem[{{Belczynski} {et~al.}(2008){Belczynski}, {Kalogera}, {Rasio}, {Taam},
  {Zezas}, {Bulik}, {Maccarone}, \& {Ivanova}}]{belc2008}
{Belczynski}, K., {Kalogera}, V., {Rasio}, F.~A., {et~al.} 2008,
  \href{http://dx.doi.org/10.1086/521026}{\JournalTitle{\apjs}, 174, 223}

\bibitem[{{Belczynski} {et~al.}(2016){Belczynski}, {Heger}, {Gladysz},
  {Ruiter}, {Woosley}, {Wiktorowicz}, {Chen}, {Bulik}, {O'Shaughnessy}, {Holz},
  {Fryer}, \& {Berti}}]{bel2016b}
{Belczynski}, K., {Heger}, A., {Gladysz}, W., {et~al.} 2016,
  \href{http://dx.doi.org/10.1051/0004-6361/201628980}{\JournalTitle{\aap},
  594, A97}

\bibitem[{Belczynski {et~al.}(2018)Belczynski, Askar, Arca-Sedda, Chruslinska,
  Donnari, Giersz, Benacquista, Spurzem, Jin, Wiktorowicz,
  {et~al.}}]{belczynski2018origin}
Belczynski, K., Askar, A., Arca-Sedda, M., {et~al.} 2018, \JournalTitle{\aap},
  615, A91

\bibitem[{{Beniamini} \& {Piran}(2016)}]{BeniaminiPiran2016}
{Beniamini}, P., \& {Piran}, T. 2016,
  \href{http://dx.doi.org/10.1093/mnras/stv2903}{\JournalTitle{\mnras}, 456,
  4089}

\bibitem[{{Broekgaarden} {et~al.}(2021){Broekgaarden}, {Berger}, {Neijssel},
  {Vigna-G{\'o}mez}, \& et~al.}]{BroekgaardenBerger2021}
{Broekgaarden}, F.~S., {Berger}, E., {Neijssel}, C.~J., {Vigna-G{\'o}mez}, A.,
  \& et~al. 2021, \JournalTitle{arXiv e-prints}, arXiv:2103.02608

\bibitem[{{Clausen} {et~al.}(2013){Clausen}, {Sigurdsson}, \&
  {Chernoff}}]{clausen2013black}
{Clausen}, D., {Sigurdsson}, S., \& {Chernoff}, D.~F. 2013,
  \href{http://dx.doi.org/10.1093/mnras/sts295}{\JournalTitle{\mnras}, 428,
  3618}

\bibitem[{{Coughlin} {et~al.}(2020){Coughlin}, {Dietrich}, {Antier}, {Bulla},
  \& et~al.}]{CoughlinDietrich2020}
{Coughlin}, M.~W., {Dietrich}, T., {Antier}, S., {Bulla}, M., \& et~al. 2020,
  \href{http://dx.doi.org/10.1093/mnras/stz3457}{\JournalTitle{\mnras}, 492,
  863}

\bibitem[{{de Mink} \& {Mandel}(2016)}]{demink2016}
{de Mink}, S.~E., \& {Mandel}, I. 2016,
  \href{http://dx.doi.org/10.1093/mnras/stw1219}{\JournalTitle{\mnras}, 460,
  3545}

\bibitem[{{Duch{\^e}ne} \& {Kraus}(2013)}]{duch2013}
{Duch{\^e}ne}, G., \& {Kraus}, A. 2013,
  \href{http://dx.doi.org/10.1146/annurev-astro-081710-102602}{\JournalTitle{Annual
  Review of Astronomy and Astrophysics}, 51, 269}

\bibitem[{{Dvorkin} {et~al.}(2015){Dvorkin}, {Silk}, {Vangioni}, {Petitjean},
  \& et~al.}]{DvorkinSilk2015}
{Dvorkin}, I., {Silk}, J., {Vangioni}, E., {Petitjean}, P., \& et~al. 2015,
  \href{http://dx.doi.org/10.1093/mnrasl/slv085}{\JournalTitle{\mnras}, 452,
  L36}

\bibitem[{{Foucart}(2012)}]{Foucart2012}
{Foucart}, F. 2012,
  \href{http://dx.doi.org/10.1103/PhysRevD.86.124007}{\JournalTitle{\prd}, 86,
  124007}

\bibitem[{{Foucart} {et~al.}(2018){Foucart}, {Hinderer}, \&
  {Nissanke}}]{FoucartHinderer2018}
{Foucart}, F., {Hinderer}, T., \& {Nissanke}, S. 2018,
  \href{http://dx.doi.org/10.1103/PhysRevD.98.081501}{\JournalTitle{\prd}, 98,
  081501}

\bibitem[{{Fragione} \& {Banerjee}(2020)}]{FragioneBanerjee2020}
{Fragione}, G., \& {Banerjee}, S. 2020,
  \href{http://dx.doi.org/10.3847/2041-8213/abb671}{\JournalTitle{\apjl}, 901,
  L16}

\bibitem[{{Fragione} \& {Loeb}(2019{\natexlab{a}})}]{frl2019a}
{Fragione}, G., \& {Loeb}, A. 2019{\natexlab{a}},
  \href{http://dx.doi.org/10.1093/mnras/stz1131}{\JournalTitle{\mnras}, 486,
  4443}

\bibitem[{{Fragione} \& {Loeb}(2019{\natexlab{b}})}]{frl2019b}
---. 2019{\natexlab{b}},
  \href{http://dx.doi.org/10.1093/mnras/stz2902}{\JournalTitle{\mnras}, 490,
  4991}

\bibitem[{{Fragione} \& {Loeb}(2021)}]{FragioneLoeb2021}
---. 2021,
  \href{http://dx.doi.org/10.1093/mnras/stab666}{\JournalTitle{\mnras}, 503,
  2861}

\bibitem[{{Fragione} {et~al.}(2018){Fragione}, {Pavl{\'\i}k}, \&
  {Banerjee}}]{frag2018}
{Fragione}, G., {Pavl{\'\i}k}, V., \& {Banerjee}, S. 2018,
  \href{http://dx.doi.org/10.1093/mnras/sty2234}{\JournalTitle{\mnras}, 480,
  4955}

\bibitem[{{Giacobbo} \& {Mapelli}(2018)}]{gm2018}
{Giacobbo}, N., \& {Mapelli}, M. 2018,
  \href{http://dx.doi.org/10.1093/mnras/sty1999}{\JournalTitle{\mnras}, 480,
  2011}

\bibitem[{{Hinderer} {et~al.}(2019){Hinderer}, {Nissanke}, {Foucart},
  {Hotokezaka}, \& et~al.}]{HindererNissanke2019}
{Hinderer}, T., {Nissanke}, S., {Foucart}, F., {Hotokezaka}, K., \& et~al.
  2019,
  \href{http://dx.doi.org/10.1103/PhysRevD.100.063021}{\JournalTitle{\prd},
  100, 063021}

\bibitem[{{Hobbs} {et~al.}(2005){Hobbs}, {Lorimer}, {Lyne}, \&
  {Kramer}}]{hobbs2005}
{Hobbs}, G., {Lorimer}, D.~R., {Lyne}, A.~G., \& {Kramer}, M. 2005,
  \href{http://dx.doi.org/10.1111/j.1365-2966.2005.09087.x}{\JournalTitle{\mnras},
  360, 974}

\bibitem[{{Hotokezaka} {et~al.}(2018){Hotokezaka}, {Beniamini}, \&
  {Piran}}]{HotokezakaBeniamini2018}
{Hotokezaka}, K., {Beniamini}, P., \& {Piran}, T. 2018,
  \href{http://dx.doi.org/10.1142/S0218271818420051}{\JournalTitle{International
  Journal of Modern Physics D}, 27, 1842005}

\bibitem[{Hurley {et~al.}(2000)Hurley, Pols, \& Tout}]{hurley2000comprehensive}
Hurley, J.~R., Pols, O.~R., \& Tout, C.~A. 2000, \JournalTitle{\mnras}, 315,
  543

\bibitem[{Hurley {et~al.}(2002)Hurley, Tout, \& Pols}]{hurley2002evolution}
Hurley, J.~R., Tout, C.~A., \& Pols, O.~R. 2002, \JournalTitle{\mnras}, 329,
  897

\bibitem[{{Janka}(2013)}]{Janka2013}
{Janka}, H.-T. 2013,
  \href{http://dx.doi.org/10.1093/mnras/stt1106}{\JournalTitle{\mnras}, 434,
  1355}

\bibitem[{{Kalogera}(2000)}]{Kalogera2000}
{Kalogera}, V. 2000,
  \href{http://dx.doi.org/10.1086/309400}{\JournalTitle{\apj}, 541, 319}

\bibitem[{{Kremer} {et~al.}(2020){Kremer}, {Ye}, {Chatterjee}, {Rodriguez}, \&
  {Rasio}}]{Kremer2019d}
{Kremer}, K., {Ye}, C.~S., {Chatterjee}, S., {Rodriguez}, C.~L., \& {Rasio},
  F.~A. \href{http://dx.doi.org/10.1017/S1743921319007269}{2020, 351, 357}

\bibitem[{{Kroupa}(2001)}]{kroupa2001}
{Kroupa}, P. 2001,
  \href{http://dx.doi.org/10.1046/j.1365-8711.2001.04022.x}{\JournalTitle{\mnras},
  322, 231}

\bibitem[{{Kruckow} {et~al.}(2018){Kruckow}, {Tauris}, {Langer}, {Kramer}, \&
  {Izzard}}]{kruc2018}
{Kruckow}, M.~U., {Tauris}, T.~M., {Langer}, N., {Kramer}, M., \& {Izzard},
  R.~G. 2018,
  \href{http://dx.doi.org/10.1093/mnras/sty2190}{\JournalTitle{\mnras}, 481,
  1908}

\bibitem[{{Kyutoku} {et~al.}(2014){Kyutoku}, {Ioka}, \&
  {Shibata}}]{KyutokuIoka2014}
{Kyutoku}, K., {Ioka}, K., \& {Shibata}, M. 2014,
  \href{http://dx.doi.org/10.1093/mnrasl/slt128}{\JournalTitle{\mnras}, 437,
  L6}

\bibitem[{{Madau} \& {Dickinson}(2014)}]{MadauDickinson2014}
{Madau}, P., \& {Dickinson}, M. 2014,
  \href{http://dx.doi.org/10.1146/annurev-astro-081811-125615}{\JournalTitle{\araa},
  52, 415}

\bibitem[{{Madau} \& {Fragos}(2017)}]{MadauFragos2017}
{Madau}, P., \& {Fragos}, T. 2017,
  \href{http://dx.doi.org/10.3847/1538-4357/aa6af9}{\JournalTitle{\apj}, 840,
  39}

\bibitem[{{Pannarale} {et~al.}(2011){Pannarale}, {Tonita}, \&
  {Rezzolla}}]{Pannarale2011}
{Pannarale}, F., {Tonita}, A., \& {Rezzolla}, L. 2011,
  \href{http://dx.doi.org/10.1088/0004-637X/727/2/95}{\JournalTitle{\apj}, 727,
  95}

\bibitem[{{Planck Collaboration} {et~al.}(2016){Planck Collaboration}, {Ade},
  {Aghanim}, {Arnaud}, \& et~al.}]{PlanckCollaborationAde2016}
{Planck Collaboration}, {Ade}, P.~A.~R., {Aghanim}, N., {Arnaud}, M., \& et~al.
  2016,
  \href{http://dx.doi.org/10.1051/0004-6361/201525830}{\JournalTitle{\aap},
  594, A13}

\bibitem[{{Podsiadlowski} {et~al.}(2004){Podsiadlowski}, {Langer},
  {Poelarends}, {Rappaport}, {Heger}, \& {Pfahl}}]{pod2004}
{Podsiadlowski}, P., {Langer}, N., {Poelarends}, A.~J.~T., {et~al.} 2004,
  \href{http://dx.doi.org/10.1086/421713}{\JournalTitle{\apj}, 612, 1044}

\bibitem[{{Raghavan} {et~al.}(2010)}]{ragh10}
{Raghavan}, D., {et~al.} 2010, \JournalTitle{\apj Suppl}, 190, 1

\bibitem[{{Rastello} {et~al.}(2020){Rastello}, {Mapelli}, {Di Carlo},
  {Giacobbo}, {Santoliquido}, {Spera}, \& {Ballone}}]{rastello2020}
{Rastello}, S., {Mapelli}, M., {Di Carlo}, U.~N., {et~al.} 2020,
  \JournalTitle{arXiv e-prints}, arXiv:2003.02277

\bibitem[{{Repetto} {et~al.}(2012){Repetto}, {Davies}, \&
  {Sigurdsson}}]{RepettoDavies2012}
{Repetto}, S., {Davies}, M.~B., \& {Sigurdsson}, S. 2012,
  \href{http://dx.doi.org/10.1111/j.1365-2966.2012.21549.x}{\JournalTitle{\mnras},
  425, 2799}

\bibitem[{{Sana}(2017)}]{Sana2017}
{Sana}, H. 2017, \href{http://dx.doi.org/10.1017/S1743921317003209}{in The
  Lives and Death-Throes of Massive Stars, ed. J.~J. {Eldridge}, J.~C. {Bray},
  L.~A.~S. {McClelland}, \& L.~{Xiao}, Vol. 329}, 110

\bibitem[{{Sana} {et~al.}(2012)}]{sana12}
{Sana}, H., {et~al.} 2012,
  \href{http://dx.doi.org/10.1126/science.1223344}{\JournalTitle{Science}, 337,
  444}

\bibitem[{{Santoliquido} {et~al.}(2020){Santoliquido}, {Mapelli}, {Bouffanais},
  {Giacobbo}, \& et~al.}]{SantoliquidoMapelli2020}
{Santoliquido}, F., {Mapelli}, M., {Bouffanais}, Y., {Giacobbo}, N., \& et~al.
  2020, \href{http://dx.doi.org/10.3847/1538-4357/ab9b78}{\JournalTitle{\apj},
  898, 152}

\bibitem[{{Shao} \& {Li}(2021)}]{ShaoLi2021}
{Shao}, Y., \& {Li}, X.-D. 2021, \JournalTitle{arXiv e-prints},
  arXiv:2107.03565

\bibitem[{{Silva} {et~al.}(2021){Silva}, {Holgado},
  {C{\'a}rdenas-Avenda{\~n}o}, \& {Yunes}}]{SilvaHolgado2021}
{Silva}, H.~O., {Holgado}, A.~M., {C{\'a}rdenas-Avenda{\~n}o}, A., \& {Yunes},
  N. 2021,
  \href{http://dx.doi.org/10.1103/PhysRevLett.126.181101}{\JournalTitle{\prl},
  126, 181101}

\bibitem[{{Tagawa} {et~al.}(2021){Tagawa}, {Kocsis}, {Haiman}, {Bartos}, \&
  et~al.}]{TagawaKocsis2021}
{Tagawa}, H., {Kocsis}, B., {Haiman}, Z., {Bartos}, I., \& et~al. 2021,
  \href{http://dx.doi.org/10.3847/1538-4357/abd555}{\JournalTitle{\apj}, 908,
  194}

\bibitem[{{Thrane} {et~al.}(2020){Thrane}, {Os{\l}owski}, \&
  {Lasky}}]{ThraneOslowski2020}
{Thrane}, E., {Os{\l}owski}, S., \& {Lasky}, P.~D. 2020,
  \href{http://dx.doi.org/10.1093/mnras/staa593}{\JournalTitle{\mnras}, 493,
  5408}

\bibitem[{{Tsang} {et~al.}(2012){Tsang}, {Read}, {Hinderer}, {Piro}, \&
  {Bondarescu}}]{tsang2012}
{Tsang}, D., {Read}, J.~S., {Hinderer}, T., {Piro}, A.~L., \& {Bondarescu}, R.
  2012,
  \href{http://dx.doi.org/10.1103/PhysRevLett.108.011102}{\JournalTitle{Phys
  Rev Lett}, 108, 011102}

\bibitem[{{Yang} {et~al.}(2020){Yang}, {Gayathri}, {Bartos}, {Haiman}, \&
  et~al.}]{YangGayathri2020}
{Yang}, Y., {Gayathri}, V., {Bartos}, I., {Haiman}, Z., \& et~al. 2020,
  \href{http://dx.doi.org/10.3847/2041-8213/abb940}{\JournalTitle{\apjl}, 901,
  L34}

\bibitem[{{Ye} {et~al.}(2020){Ye}, {Fong}, {Kremer}, {Rodriguez}, {Chatterjee},
  {Fragione}, \& {Rasio}}]{ye2020}
{Ye}, C.~S., {Fong}, W.-f., {Kremer}, K., {et~al.} 2020,
  \href{http://dx.doi.org/10.3847/2041-8213/ab5dc5}{\JournalTitle{\apjl}, 888,
  L10}

\bibitem[{Ye {et~al.}(2019)Ye, Kremer, Chatterjee, Rodriguez, \&
  Rasio}]{ye2019millisecond}
Ye, C.~S., Kremer, K., Chatterjee, S., Rodriguez, C.~L., \& Rasio, F.~A. 2019,
  \JournalTitle{\apj}, 877, 122

\end{thebibliography}

\end{document}